\documentclass[preprint,aps,prd,groupedaddress,showpacs,floatfix,%
nofootinbib]{revtex4}

\usepackage{dcolumn}

\usepackage{graphicx}
\usepackage{epsfig}
\usepackage{hyperref}
\usepackage{graphicx}
\usepackage{dcolumn}

\usepackage{longtable}
\begin{document}
\title{Improved soft-wall model with a negative dilaton}
\author{Fen
Zuo\footnote{Email: zuof@ihep.ac.cn}}\affiliation{Institute of High
Energy Physics, Chinese Academy of Sciences, Beijing 100049, China}
\affiliation{Theoretical Physics Center for Science Facilities,
Chinese Academy of Sciences, Beijing 100049, China}
 \affiliation{Key Laboratory of Frontiers in Theoretical Physics,~Institute of
Theoretical Physics,~Chinese Academy of Sciences, Beijing 100190,
China}
\begin{abstract}
I propose to change the sign of the dilaton in the infrared~(IR)
soft-wall AdS/QCD model, in order to implement confinement. The
deformed model exhibits interesting properties, especially in
describing chiral symmetry breaking. The expectation value of the
scalar field $X$, which determines the quark mass and condensate,
approaches a constant in the IR limit, rather than blows up in the
original model. In contrast to the estimate in Ref.
\cite{Shifman:2007xn}, this kind of solution will not lead to chiral
symmetry restoration for highly-excited states, due to the property
of the harmonic-oscillator equation. Instead, it will guarantee the
Regge behavior of the axial meson spectrum, and also the
pseudoscalar mesons. The value of the condensate can be fixed by
requiring that the pion be massless in the chiral limit, but only
under some approximation in the present model. We also find that, by
relaxing the IR boundary conditions, the unphysical massless state
in the vector channel can be eliminated.

\end{abstract}
\keywords{QCD, AdS-CFT Correspondence}
\pacs{11.25.Tq, 
11.10.Kk, 
11.15.Tk  
12.38.Lg  
}

 \maketitle

\section{Motivation}
The holographic approach towards QCD, or AdS/QCD, has made
significant progress in describing low energy hadron dynamics recent
years. Except for predicting various low energy parameters with an
unexpected precision, it joins together many traditional approaches
to describe strong interactions, such as nonlinear chiral symmetry
realization, hidden local symmetry, QCD sum rule, light-cone
perturbative theory, etc. However, it also suffers from many
problems, one of which is the description of the experimentally well
established Regge trajectories of hadron spectrum. To overcome this,
people propose to select a special background, namely a
quadratically increasing dilaton together with a five dimensional
anti de Sitter~(AdS) space-time \cite{Karch2006}. Fields propagating
in this background satisfy harmonic-oscillator type equations, and
give exactly Regge trajectories for various mesons. The model based
on this background is often called soft-wall model, to be compared
with the hard-wall model \cite{Erlich:2005qh}, where the background
is a slice of AdS space-time.

While Regge behavior is well described in this model, we now run
into new troubles when trying to incorporating chiral symmetry
breaking. If we introduce a tachyonlike scalar field $X$ to break
chiral symmetry as in the hard-wall model, the solution for the
expectation value of $X$ now diverges exponentially in the
infrared~(IR) limit. This is unreasonable, since every physical mode
should be regular in the bulk region according to the standard
AdS/CFT prescription \cite{Witten1998a}. Practically, this solution
adds an exponentially increasing term to the equation for the axial
modes, and the previously obtained Regge behavior was ruined. There
has been some attempt to overcome this problem by adding higher
dimensional terms to the potential of $X$ \cite{Gherghetta:2009ac},
see also \cite{Sui:2009xe}. A simpler way is to change the sign of
the dilaton, and push the divergence into the background. This may
seem a little rude, but it does regularize the divergent solution.
Meantime, it gives the same Regge trajectories for the vector
mesons, which has already been pointed out in Ref.~\cite{Karch2006}.
Moreover, explicit calculations show that various mesons, including
axial and pseudoscalar ones, will fall on Regge trajectories with
the same slope.

Another motivation comes from the consideration of confinement
property of the background. The original soft-wall background is in
some sense equivalent to adding a Gaussian factor $e^{-z^2}$ to the
AdS metric. In Ref. \cite{Andreev:2006ct} an alternative deformation
factor $e^{z^2}$ was shown to result confining potential between
heavy quarks. It is just this increasing exponential which
guarantees the confining property, according to the general
criterion in Ref. \cite{Kinar:1998vq}. Therefore, in order to obtain
linear Regge trajectories in a confining background, one is
naturally led to consider a negative dilaton $\Phi\sim-z^2$. Since
the dilaton contributes to the action in the form $e^{-\Phi}$, which
now increases quickly in the IR region, this kind of background
includes actually a quasi-hard wall, not a soft wall any more.

The paper is organized as follows. In the next section, I show that
in the modified background chiral symmetry breaking can be naturally
incorporated. The Regge trajectories of various mesons, including
the axial and pseudoscalar mesons, can be reproduced in the improved
model. Some comments on the confining property of the background is
given in Sec.{I}{I}{I}. In the last section I give a short summary.

\section{Improved soft-wall Model}
The five dimensional AdS space-time can be given in the
Poincar$\acute{e}$ coordinate as:
\begin{equation}
g_{MN}dx^Mdx^N = \frac{1}{z^2}\left(\eta_{\mu \nu}dx^{\mu}dx^{\nu} -
dz^2\right) \ ,
\end{equation}
where $ \eta_{\mu\nu} = {\rm Diag}\, (1,-1,-1,-1) $ and $ \mu, \nu =
(0,1,2,3) $, $ M, N = (0,1,2,3,z) $. The dilaton is now chosen to be
\begin{equation}
\Phi(z)=-\Lambda^2z^2,
\end{equation}
which has an opposite sign to that in the original soft-wall model
\cite{Karch2006}. This kind of background was shown to exist in some
tachyon-dilaton-gravity system \cite{Csaki:2006ji}, where the close
string tachyon plays the role of the dilaton field here. The
nontrivial dilaton profile violates scale invariance explicitly, and
introduces a scale $\Lambda$ into the boundary theory. We choose to
set $\Lambda$ to be 1 and recover it from dimensional analysis when
needed. As in Ref. \cite{Erlich:2005qh}, one introduces two sets of
gauge fields, $A_{\rm{L}}$ and $A_{\rm{R}}$, to represent the gauged
chiral symmetry. Then this symmetry is broken both explicitly and
spontaneously by employing a tachyonic scalar field $X$. With the
dilaton turned on, the action reads:
\begin{equation}
\label{AdS} S =  \int d^5x \sqrt{g}~e^{-\Phi}
 ~{\rm Tr}~\left[(D^{M}X)^{\dagger}(D_{M}X) +
3 X^{\dagger}X -
\frac{1}{4g_5^2}(F_{\rm{L}}^{MN}F_{{\rm{L}}MN}+F_{\rm{R}}^{MN}F_{{\rm{R}}
MN})\right] \  ,
\end{equation}
where $ A=A^at^a, F_{MN}=\partial_M A_N-\partial_N A_M-i[A_M,A_N], D
X = \partial X - iA_{\rm{L}}X + iX A_{\rm{R}} $. The generators of
the gauge group are normalized as $\mbox{Tr}~t^at^b=\delta^{ab}/2$.

\subsection{Vacuum Solution}
A nonzero expectation value $X_0$ of $X$ will breaks the
$\rm{SU}(N_f)_L\times\rm{SU}(N_f)_R$ symmetry to the vector part. So
one first takes $A_{\rm{L}}$ and $A_{\rm{R}}$ to be zero and solves
the equation of motion for $X$. Denote the $v(z)=2X_0$, then $v(z)$
satisfies the following equation:
\begin{equation}
\partial_z\Big(\frac{1}{z^3}\,e^{-\Phi(z)}\partial_zv(z)\Big)+3\frac{1}{z^5}\,e^{-\Phi(z)}v(z)=0
\label{cond}
\end{equation}
The solution is given by
\begin{equation}
v(z)=C_1~v_1(z)+C_2~v_2(z),
\end{equation}
with $v_1(z)=ze^{-z^2}~U(-1/2,0;z^2)$ and $v_2=z^3e^{-z^2}
~_1F_1(1/2,2;z^2)$. Here $U$ is the Tricomi function, and $_1F_1$
the Kummer function. Since $U(-1/2,0;z^2)\to 1/\sqrt{\pi}$ and
$_1F_1$ goes to unity as $z\to 0$, one can identify $C_1/\sqrt{\pi}$
as the quark mass and $C_2$ the condensate so that $v(z)\sim m_q z+
\sigma z^3$. When $z\to \infty$, $v_1(z)$ vanishes as $z^2e^{-z^2}$
and $v_2(z)\to 1/\sqrt{\pi}$, while in the original model one
solution goes to a constant and the other diverges like
$z^2e^{z^2}$. So the divergence of $v(z)$ is successfully absorbed
into the background, and now $v(z)$ becomes regular in the bulk.
This will be important in guaranteeing the Regge trajectories of the
axial and pseudoscalar mesons, as shown in the following.

\subsection{Vector Sector}
Now consider the fluctuations around the vacuum, which will
correspond to the bound states of various mesons. First we consider
the fluctuations from the vector combination
$V=(A_{\rm{L}}+A_{\rm{R}})/2$. Taking the axial gauge $V_z=0$ and
performing Fourier transformation in the $x$ direction, the
transverse components $V_\mu^{\rm{T}}$ then satisfy the following
equation:
\begin{equation}
\partial_z(e^{-B}\partial_zV^{\rm{T}}_\mu(q,z))+q^2e^{-B}V^{\rm{T}}_\mu(q,z)=0,
\end{equation}
where $B=\Phi(z)+\log z$. The normalizable solutions $v_n$, obtained
at $q^2=m_{v_n}^2$, are then considered to be dual to vector mesons.
Applying a Bogoliubov transformation $v_n=e^{B/2}\psi^v_n$, one
obtains a Schr$\ddot{o}$dinger equation:
\begin{equation}
-{\psi^v_n}''+V_v(z)\psi^v_n=m_{v_n}^2\psi^v_n,\label{eq:V}
\end{equation}
with the potential
\begin{equation}
V_v(z)=(B')^2/4-B{''}/2=z^2+3/(4z^2).
\end{equation}
Now the equation for $\psi^v_n$ is exactly the same as in the
original soft-wall model, and the regular solutions are:
\begin{equation}
\psi^v_n(z)=e^{-z^2/2}z^{3/2}\sqrt{\frac{2}{n+1}}L^1_n(z^2),
\end{equation}
with $L^m_n$ the associated Laguerre polynomials and
\begin{equation}
m^2_{v_n}=4(n+1),~n=0,1,2,...
\end{equation}
One can then use the experimental value for $m_\rho$ to determine
the parameter $\Lambda$, which gives $\Lambda=m_\rho/2\simeq 0.388
\mbox{GeV}$. The form of corresponding gauge fields reads
\begin{equation}
v_n(z)=e^{-z^2}z^2\sqrt{\frac{2}{n+1}}L^1_n(z^2).\label{eq:rho}
\end{equation}
An extra exponential $e^{-z^2}$ now appears in $v_n$ which forces it
to vanish at the IR boundary, while in Ref. \cite{Karch2006} $v_n$
diverges as $z^{2n+2}$. The normalization condition is of the same
form:
\begin{equation}
\int_0^\infty\frac{e^{-\Phi}}{z}~v_mv_n~dz=\delta_{mn}.
\end{equation}

One can also find the non-normalizable solution to Eq. (\ref{eq:V})
at spacelike momentum $q^2<0$. This kind of solution can be
separated into the source $V_\mu^0(q)$ and the bulk-to-boundary
propagator $V(q,z)$ as $V^{\rm{T}}_\mu(q,z)=V(q,z)V_\mu^0(q)$.  The
general solution for $V(q,z)$ 
is given by
\begin{equation}
V(q,z)=A~e^{-z^2}~U(\frac{-q^2}{4},0,z^2)+B~z^2~e^{-z^2}~_1F_1(1-\frac{q^2}{4},2,z^2).\label{eq:Vqz}
\end{equation}
At large $z$, $U(\frac{-q^2}{4},0,z^2)\sim~(z^2)^{q^2/4}$ and
$_1F_1(1-\frac{q^2}{4},2,z^2)\sim e^{z^2}(z^2)^{-1-q^2/4}$. When
$q^2<0$, the second part of the solution diverges
as$z\rightarrow\infty$, and thus should be discarded according to
the standard {``}regularity in the bulk"
condition~\cite{Witten1998a}. Then $A$ can be determined from the
boundary condition $V(q,0)=1$ to be $A=\Gamma(1-q^2/4)$. From the
solution of $V(q,z)$, one can derive the two-point function, which
turns out to be
\begin{eqnarray}
  \Pi_{\rm V}(q^2) \!&=&\!
\left.\frac{e^{z^2}~V(q,z)}{g_5^2q^2} \frac{\partial_z
V(q,z)}{z}\right|_{z=\epsilon}\nonumber\\
\!&=&\!
-\frac{1}{2g_5^2}\psi\left(1-\frac{q^2}{4}\right)-\frac{2}{g_5^2q^2}+\rm{Constant}.\label{eq:OPE2}
\end{eqnarray}
Comparing to our previous result \cite{Zuo:2008re}, an additional
term $-\frac{2}{g_5^2q^2}$ appears, which seems to correspond to a
{``}pion" pole in the vector mode. This unphysical mode has already
been pointed out in the pioneer work of Karch et
al~\cite{Karch2006}. It also affects the three point functions,
e.g., the electromagnetic~(EM) form factors of the vector mesons.
Following the derivation in Ref. \cite{Radyushkin2007}, one find
that the EM form factors are given by
\begin{equation}\label{formfac}
F_{nk}(q^2)  = \int^{\infty}_{0} \frac{dz}{z} \, e^{z^2} V(q,z) \,
v_n(z)  \,  v_k(z)   \ .
\end{equation}
At $q^2=0$ the off-diagonal transitions do not vanish and the
diagonal ones do not normalize to unity, which signal a violation of
charge conservation.

The underlying reason for all these is that, $V(0,z)=e^{-z^2}$ is
not a flat function as expected. One may ask, if we can relax the IR
boundary condition\footnote{Such kind of modification has once been
investigated in Ref.~\cite{Colangelo:2007if}.} to make $V(0,z)$
flat, at least approximately near the ultraviolet~(UV) boundary?
This sounds reasonable, since $V(q,z)$ does not describe a physical
state, or the physical vacuum. To this end, we retain the IR
divergent part in the general solution (\ref{eq:Vqz}) with $B$ a
general function of $q^2$. Since this part is sub-leading in the UV
region, it does not affect the determination of the coefficient $A$.
Now we calculate the correlation function again. In general there
will be additional contributions from the IR boundary, since now
neither $V(q,z)$ or its derivative vanishes there. They are unwanted
from the holographic point of view, and can be eliminated by adding
suitable boundary terms. Then the correlation function can be
derived to be:
\begin{eqnarray}
  \Pi_{\rm V}(q^2) \!&=&\!
\left.\frac{e^{z^2}~V(q,z)}{g_5^2q^2} \frac{\partial_z
V(q,z)}{z}\right|_{z=\epsilon}\nonumber\\
\!&=&\!
-\frac{1}{2g_5^2}\psi\left(1-\frac{q^2}{4}\right)+\frac{2(B(q^2)-1)}{g_5^2q^2}+\rm{Constant}.\label{eq:OPE3}
\end{eqnarray}
So the IR divergent part contributes to the $q^2=0$ pole too. To
eliminate this unphysical pole, we are forced to choose $B(0)=1$.
Interestingly, with this choice $V(0,z)$ becomes
\begin{equation}
V(0,z)=e^{-z^2}+z^2~e^{-z^2}~_1F_1(1,2,z^2)=1,
\end{equation}
and charge conservation is restored. Since now $V(q,z)$ does not
vanish at the IR boundary as the normalizable modes do, the
decomposition formula~\cite{Strassler2004} of $V(q,z)$ does not
exist any more. As a result, general vector meson dominance will be
violated, and there will be direct couplings of the external fields
to the mesons. The complete form of $V(q,z)$ and the
phenomenological effects of it are still under investigation.

\subsection{Axial Sector}
The fluctuations of the axial combination
$A=(A_{\rm{L}}-A_{\rm{R}})/2$ are a little complicated. The axial
field can be separated into the transverse and longitudinal parts as
$A_\mu=A_{\mu\perp}+\partial_\mu \varphi$, where $\partial^\mu
A_{\mu\perp}=0$. The longitudinal part $\varphi$ will get entangled
with the pseudoscalar fluctuations, so we decay to talk about them
in the next subsection. The transverse part, after taking the axial
gauge $A_z=0$, satisfies the equations:

\renewcommand\d{\partial}

\begin{equation}
  \left[ \d_z\left(\frac{e^{-\Phi}}{z} \, \d_z A^a_\mu \right) +
\frac{q^2 e^{-\Phi}}{z}
  A^a_\mu - \frac{g_5^2 \, v(z)^2 e^{-\Phi}}{z^3} A^a_\mu\right]_\perp =0 \, ,
\label{eq:A}
\end{equation}
Again one takes the Bogoliubov transformation to change them into
the Schr$\ddot{o}$dinger form, and obtain the potential:
\begin{equation}
V_a(z)=z^2+\frac{3}{4z^2}+\frac{g_5^2v(z)^2}{z^2}.
\end{equation}
From the IR behavior of $v(z)$, the potential $V_a(z)$ can be
expanded as:
\begin{eqnarray}
V_a(z)&=&z^2+\frac{3}{4z^2}+\frac{g_5^2v(z)^2}{z^2}\nonumber\\\
      &=&z^2+\left[\frac{3}{4}+\frac{g_5^2C_2^2}{\pi}\right]~z^{-2}+\mbox{O}(z^{-4}).
\end{eqnarray}
Then we get the following spectrum approximately:
\begin{equation}
m_{a_n}^2\approx 4n+2\sqrt{1+\frac{g_5^2C_2^2}{\pi}}+2 \label{eq:Am}
\end{equation}
Notice that due to the nonzero quark condensate $C_2$, the vector
and axial spectrum are now split, with the leading order mass
splitting independent of the excitation number $n$. This is in
contrast to the estimation in Ref. \cite{Shifman:2007xn}, where a
constant behavior of $X$ at large $z$ was believed to introduce only
a $1/n$ deformation of the axial masses and lead to chiral symmetry
restoration for highly-excited states. The asymptotic wave functions
in the IR region are
\begin{equation}
a_n(z)=e^{-z^2}z^{m+1}\sqrt{\frac{2}{n+m}}L^m_n(z^2),\label{eq:a}
\end{equation}
where $m=\sqrt{1+\frac{g_5^2C_2^2}{\pi}}$. These solutions can not
be extended naively to the small-$z$ region since they have the
wrong $z\to 0$ behavior for general $m$. To obtain the explicit
solutions in the whole region one must employ numerical methods. It
is also interesting to study the non-normalizable mode. Because of
the complicated form of $v(z)$, this is more involved than the
vector one and we leave it to future work.

\subsection{(Pseudo)Scalar sector}
Now we turn to the fluctuations around the scalar field $X$, and
parameterizes $X$ as $X=(X_0+S)\rm{exp}(i2\pi)$ as in Ref.
\cite{DaRold:2005zs} and Ref. \cite{Colangelo:2008us}. The equation
for $S$ is easily derived \cite{Colangelo:2008us}:
\begin{equation}\label{eqscalar}
\partial_z\left(\frac{1}{z^3}\,e^{-\Phi(z)}\partial_z~S^a\right)+3\frac{1}{z^5}\,e^{-\Phi(z)}~S^a-q^2\frac{1}{z^3}\,e^{-\Phi(z)}~S^a=0\,\,\,.
\end{equation}
The Schr$\ddot{o}$dinger potential becomes $V_s(z)=z^2+\frac{3}{4
z^2}-2$, and the corresponding spectrum is given by
$m_{s_n}^2=4n+2$. Comparing with the corresponding spectrum in Ref.
\cite{Colangelo:2008us}, one see that the deformation of the dilaton
profile decreases the masses of the scalar mesons.

Finally, the pseudoscalar fluctuations $\pi$ and the longitudinal
part of the axial field satisfy the coupled equations

\renewcommand\d{\partial}
\begin{equation}
  \d_z\left(\frac{e^{-\Phi}}{z} \, \d_z \varphi^a \right)
+\frac{g_5^2 \, v(z)^2 e^{-\Phi}}{z^3} (\pi^a-\varphi^a) = 0 \, ,
\label{AL}
\end{equation}
\begin{equation}
  -q^2\d_z\varphi^a+\frac{g_5^2 \, v(z)^2}{z^2} \, \d_z \pi^a =0 \, ,
\label{Az}
\end{equation}
These two equations can be combined \cite{Kwee:2007dd} to give the
decoupled form equation:
\begin{equation} \label{decoupled}
\d_z \left[ \Gamma(z) \, \d_x y^a \right] + \Gamma(z) \left[ q^2 -
\beta(z) \right] y^a = 0 \, ,
\end{equation}
where $y^a \! \equiv \! [e^{-\Phi(z)}/z][\d_z \varphi ^a]$,
$\beta(z) \! \equiv g_5^2 \, v(z)^2 / z^2$ and $\Gamma(z) \! \equiv
z/[\beta(z) e^{-\Phi(z)}]$. One can then change this equation into
the Schr$\ddot{o}$dinger form with the Bogoliubov transformation
$y=e^{\tilde{B}/2}\tilde y$ with
\begin{equation}
\tilde{B}(z)=2\log v(z)-\Phi(z)-3\log z.
\end{equation}
The corresponding Schr$\ddot{o}$dinger potential becomes:
\begin{equation}
V_p=\frac{{\tilde B}'^{2}}{4}-\frac{{\tilde
B}{''}}{2}+\frac{g_5^2~v(z)^2}{z^2}.
\end{equation}
By using the asymptotic behavior of $v(z)$ at large $z$, the
potential can be expanded as:
\begin{equation}
V_p=z^2+(\frac{g_5^2C_2^2}{\pi}-\frac{9}{4})~z^{-2}-4+\mbox{O}(z^{-4})
\end{equation}
Then the spectrum is approximately given by
\begin{equation}
m_{p_n}^2\approx~4n+2~\sqrt{\frac{g_5^2C_2^2}{\pi}-2}-2 .
\end{equation}
Since the quark mass $C_1$ does not appear, the lowest state must be
massless. To satisfy this condition, one must choose
$\frac{g_5^2C_2^2}{\pi}=3$ at this order, which then induces that
$m_{p_n}^2\approx~4n$ and $m_{a_n}^2\approx~4n+6$. However, when one
further expand the potential to include higher powers of $z^{-1}$,
variation of $C_2$ is not enough to ensure that the lowest state be
a massless state. This has been confirmed by numerical calculation.
It is interesting to study whether these higher powers can be
eliminated by fine-tuning the detailed form of $\Phi(z)$ and the
metric in the IR region. If this is true, we can always determine
the quark condensate by requiring the lowest state being massless. A
physical $\pi$ mass can then be obtained by taking the quark mass
term as perturbation. Since all the normalizable modes are
suppressed in the IR region as $z^ne^{-z^2}$, one can easily derive
the Gell-Mann-Oakes-Renner (GOR) relation following the procedure of
Refs.~\cite{Abidin:2009aj,Zuo:2009hz}.

\section{Comments on the confinement property of the model}
To explore the confining property of the background, one could
compute the quark-antiquark potential through the prescription of
Refs. \cite{Maldacena:1998im,Rey:1998ik}. Namely one choose a
rectangle loop $C$ with one time direction, and calculate the
expectation value of the Wilson loop from the proper area of a
fundamental string worldsheet ending on $C$ at the boundary. When a
nontrivial dilaton is turned on, the generalization is not so
straightforward. One way to implement the dilaton was to work in the
string frame rather than Einstein frame metric, as done in Ref.
\cite{Gursoy:2007er}. However, there are still some ambiguities in
doing so due to the complicated worldsheet coupling to the
background dilaton field~\footnote{Thanks Oleg Andreev for pointing
out this to me.}. To avoid this ambiguity, we choose to consider
some phenomenologically equivalent background in which the dilaton
is trivial. In the vector sector, the soft-wall model can also be
described by adding a warp factor to the AdS metric
\cite{Andreev2006}:
\begin{equation}
g_{MN}dx^Mdx^N = \frac{A(z)}{z^2}\left(\eta_{\mu
\nu}dx^{\mu}dx^{\nu} - dz^2\right) \ ,
\end{equation}
with $A(z)=e^{-2\Phi}$. Such kind of metric has been studied first
in Ref. \cite{Ghoroku:2005vt}, but only as a deformation of the
hard-wall model. To calculate the quark-antiquark potential, one
first defines the following functions:
\begin{eqnarray}
f^2(z)&=&-g_{00} g_{\mu\mu} \nonumber\\
g^2(z)&=&-g_{00} g_{zz}
\end{eqnarray}
Generally, the string worldsheet will not pass the point $z_m$ where
$f(z)$ has a minimum or $g(z)$ diverges. When the distance $L$
between the quark and antiquark is large enough, most of the
worldsheet lies close to $z_m$. This part gives the dominant
contribution to the potential, $E\approx f(z_m)L$. Thus confinement
occurs if and only $f(z_m)\neq 0$, and the corresponding confining
tension is given by $f(z_m)$~\cite{Kinar:1998vq}. For the above
metric, one has $f(z)=g(z)=e^{-2\Phi}/z^2$. The criterion can only
be fulfilled when $\Phi(z)$ goes to $-\infty$ as
$z\rightarrow\infty$. From this one easily concludes that the
original background with $\Phi(z)\sim z^2$ in Ref. \cite{Karch2006}
is not confining. For the present choice $\Phi(z)=-z^2$, $f(z)$ has
a nonzero minimum $f(z_m)=2\mbox{e}$ at $z_m=\frac{\sqrt{2}}{2}$,
leading to a nonvanishing confining potential. The detailed
calculation of the quark potential in such a background confirms the
above estimation~\cite{Andreev:2006ct}. At finite temperature, the
deconfinement phase transition indeed happens and exhibits
interesting properties, see, e.g.,
\cite{Andreev:2006eh,Colangelo:2010pe}.

 \section{Summary}
The properties of a {``}soft-wall" model with a negative dilaton are
discussed. All the physical fluctuations become regular in the
infrared region after the modification. The vacuum solution of the
field $X$, which is dual to the quark bilinear operator, tends to a
constant in the IR limit. This ensures the Regge behavior of the
spectrum of the pseudoscalar and axial mesons. At the same time,
this does not lead to chiral symmetry restoration for highly-excited
mesons. Under some assumptions, the value of the quark condensate
can be determined by requiring the pion being massless in the chiral
limit. We also point out that the present choice for the dilaton
leads to confinement, while that of the original soft-wall model
does not. The unphysical pion mode in the vector channel seems to
result from the improper IR boundary condition. Once the boundary
condition is relaxed, the unphysical state can be eliminated.

\vskip2cm

{\bf Note Added:}  While writing this paper, it was noted that
Ref.~\cite{deTeramond:2009xk} also proposes to modify the sign of
the dilaton in the soft-wall model to obtain confinement.

\section{Acknowledgements}
\label{sec:acknowledge}

I would like to thank Yi-hong Gao for very useful discussions, and
Andreas Karch, Stanley Brodsky, Guy F. de Teramond, Elias Kiritsis,
Oleg Andreev, Sergey Afonin for their helpful comments and
correspondences.

\newpage

\end{document}